\documentstyle[twocolumn]{jpsj2}

\def\runtitle{Lattice Distortion and Resonant X-Ray Scattering in DyB$_{2}$C$_2$}
\def\runauthor{Jun-ichi {\sc Igarashi} and Tatsuya {\sc Nagao}}

\title{%
Lattice Distortion and Resonant X-Ray Scattering in DyB$_{2}$C$_2$}

\author{%
    Jun-ichi {\sc Igarashi}\footnote{E-mail: jigarash@spring8.or.jp}
    and Tatsuya {\sc Nagao}$^{1}$}

\inst{%
Synchrotron Radiation Research Center, Japan Atomic Energy Research Institute, 
Mikazuki, Sayo, Hyogo  679-5148 \\
$^{1}$Faculty of Engineering, Gunma University, Kiryu, Gunma 376-8515
}

\recdate{ January 6, 2003 }

\abst{%
We study the resonant x-ray  scattering (RXS) spectra
at the Dy $L_{\rm III}$ absorption edge in the quadrupole ordering phase
of DyB$_2$C$_2$.
Analyzing the buckling of sheets of B and C atoms, 
we construct an effective model 
that the crystal field is acting on the $5d$ and $4f$ states with the 
principal axes different for different sublattices.
Treating the $5d$ states as a band and the $4f$ states as localized states,
we calculate the spectra within the dipole transition.
We take account of processes that (1) the lattice distortion directly
modulates the $5d$ states and (2) the charge anisotropy of the quadrupole
ordering $4f$ states modulates the $5d$ states through the $5d$-$4f$
Coulomb interaction.
Both processes give rise to the RXS intensities on 
$(00\frac{\ell}{2})$ and $(h0\frac{\ell}{2})$ spots.
Both give similar photon-energy dependences and the same 
azimuthal-angle dependences for the main peak,
 in agreement with the experiment. 
The first process is found to give the intensities much larger than the second
one in a wide parameter range of crystal field.
This suggests that the main-peak of the RXS spectra is not a direct
reflection of the quadrupole order but mainly controlled 
by the lattice distortion.
}

\kword{%
resonant X-ray scattering, DyB$_{2}$C$_{2}$, 
       lattice distortion, Dy $L_{{\rm III}}$ absorption edge}

\begin{document}
\sloppy
\maketitle
\def\ncr{\nonumber\\}

\newpage
\section{Introduction}
Resonant x-ray scattering has recently attracted much interest,
since the resonant enhancement for the prohibited Bragg reflection
corresponding to the orbital order has been observed in several
transition-metal compounds by using synchrotron radiation with photon energy 
around the $K$ absorption edge.
\cite{Murakami98a,Murakami98b,Murakami99c,Murakami00b}
For such $K$-edge resonances, $4p$ states of transition metals are involved 
in the intermediate state in the electric dipolar ($E_1$) process, 
and they have to be modulated in accordance with the orbital order 
for the signal to be observed.
This modulation was first considered to come from
the anisotropic term of the $4p$-$3d$ intra-atomic Coulomb interaction,
\cite{Ishihara1} 
but subsequent studies based on the band structure calculation
\cite{Elfimov,Benfatto,Takahashi1,Takahashi2} have revealed that
the modulation comes mainly from the crystal distortion via 
the oxygen potential on the neighboring sites.
This is because $4p$ states are so extending in space 
that they are very sensitive to the electronic structure at neighboring sites. 

Rare-earth compounds also
show the orbital order (usually an ordering of quadrupole moments). 
In CeB$_6$, RXS experiments were carried out 
around the Ce $L_{\rm III}$ absorption edge, and resonant enhancements
have been found  on quadrupolar ordering superlattice spots.
\cite{Nakao01} Only one peak appeared as a function
of photon energy, which was assigned to the $E_1$ process.
In the $E_1$ process, $5d$ states of Ce in the intermediate state
are to be modulated in accordance with the superlattice spots.
Since the lattice distortion seems extremely small and the $5d$ states are
less extending than the $4p$ states in transition-metal compounds,
it is highly possible that the modulation is mainly caused by the Coulomb 
interaction between the $5d$ states and the orbital ordering $4f$ states.
In our previous papers,\cite{Nagao,Igarashi} we demonstrated this scenario by
calculating the RXS spectra on the basis of the effective Hamiltonian 
of Shiina et al.\cite{Shiina,Sakai,Shiba}
Without the help of lattice distortion, 
we obtained sufficient intensities of the 
spectra, and reproduced well the temperature and magnetic field dependences.
This situation contrasts with those in transition-metal compounds.

\begin{figure}[t]
\includegraphics[width=8.0cm]{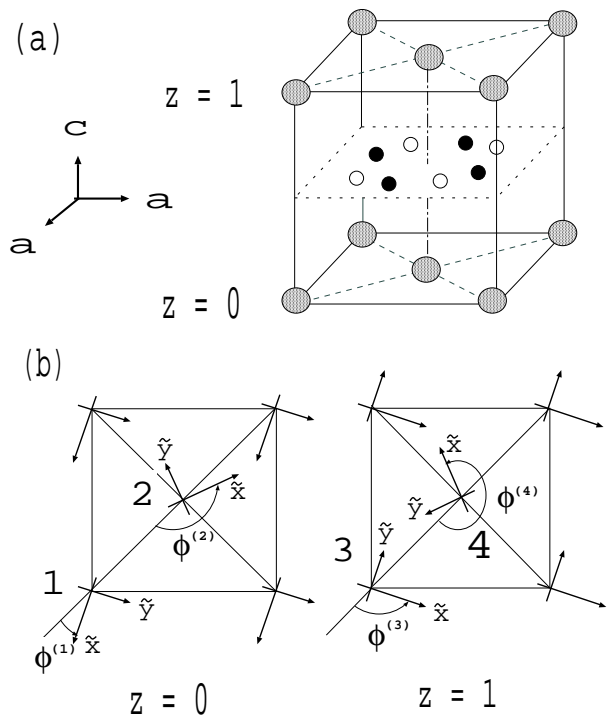}
\caption{
(a) Sketch of the crystal structure of DyB$_2$C$_2$
($P4/mbm$: $a=5.341$ ${\rm \AA}$, $c=3.547$ ${\rm \AA}$ at $30$ K).
Gray large circles are Dy atoms. Solid and open small circles 
are B and C atoms, respectively. 
(b) Local coordinate frames attached to each sublattice.
\label{fig.cryst}}
\end{figure}

Another example for rare-earth compounds is RXS experiments 
on DyB$_2$C$_2$, where the intensity 
is resonantly enhanced near the Dy $L_{\rm III}$ absorption edge.
\cite{Tanaka,Hirota,Matsumura}
This material takes a tetragonal form at high temperatures as shown
in Fig. \ref{fig.cryst}(a), and 
undergoes two phase transitions with decreasing temperatures
in the absence of the magnetic field:
a quadrupole order below $T_{\rm Q}$ ($=24.7$ K) 
(Phase II) and a magnetic order below $T_{\rm C}$ ($=15.3$ K) 
(Phase III).\cite{Yamauchi}
Corresponding to the transition at $T_{\rm Q}$, 
a large non-resonant intensity is found in the $\sigma\to\sigma'$ channel 
on the $(h0\frac{\ell}{2})$ spot 
($h$ and $\ell$ are odd integers).\cite{Matsumura}
This suggests that some structural change takes place at $T=T_{\rm Q}$
from the tetragonal phase at high temperatures.\cite{Tanaka,Hirota} 
A buckling of sheets of B and C atoms was proposed,\cite{Tanaka} 
and the non-resonant intensities by the buckling
has recently been evaluated;
about $0.01$ ${\rm \AA}$ shift of B and/or C 
atoms may be sufficient to give rise to
such large intensities.\cite{Adachi}
It is not clear in experiments whether the intensity on this spot 
is resonantly enhanced at the $L_{\rm III}$ edge,
since the non-resonant part is so large that it may mask the resonant behavior.
On the other hand, the resonant enhancement of RXS intensities has clearly
been observed on the superlattice spot $(00\frac{\ell}{2})$.

In this paper, we study the mechanism of the RXS spectra at the $L_{\rm III}$ 
edge in Phase II of DyB$_2$C$_2$.
Since the $5d$ states are so extended in space that
they are sensitive to lattice distortion caused by the buckling of 
sheets of B and C 
atoms. Then the question arises whether the direct influence of 
the lattice distortion on the $5d$ states is larger than the influence of the 
anisotropic $4f$ charge distribution associated with the quadrupole order
through the $5d$-$4f$ Coulomb interaction. 
Lovesey and Knight\cite{Lovesey} have discussed the mechanism from the
symmetry viewpoint, and have pointed out that the RXS intensities 
on $(00\frac{\ell}{2})$ and $(h0\frac{\ell}{2})$ spots come from lowering 
the local symmetry probably due to lattice distortion. 
The argument based on symmetry alone is powerful in some respect,
but does not shed light on this issue.
In the transition-metal compounds, the corresponding question 
has already been answered 
by {\emph ab initio} calculations as mentioned above. 
However, such {\emph ab initio} calculations are difficult in rare-earth 
compounds. We resort to a model calculation by
treating the $5d$ states as a band and the $4f$ states as localized states.
The buckling of sheets of B and C atoms causes modulations of the $5d$ bands and 
of the $4f$ states. We analyze such effects of lattice distortion 
on the basis of the point charge model,\cite{Hutchings}
which leads to four inequivalent 
Dy sites with principal axes of the crystal field shown 
in Fig.~\ref{fig.cryst}(b). 
These principal axes seem to correspond well to the direction of
magnetic moments in the magnetic phase.\cite{Yamauchi}
Of course, the point charge model is not good in quantitative viewpoint.
Nonetheless, we construct an effective model that the $5d$ and $4f$ states
are under the crystal field of the same form and with the same principal axes 
as the above analysis.

The crystal field modulates the $5d$ states.
Although the actual effect may come from hybridizations to
$2p$, $3s$ states of B and C, it can be included into a form 
of the crystal field.
The crystal field also makes the quadrupole moment of the $4f$ states align
along the principal axes, establishing a quadrupole order.
A molecular field caused by the Dy-Dy interaction may also act on the $4f$
states in Phase II in addition to the crystal field.
This interaction may be mediated by the RKKY interaction, 
but the explicit form has not been derived yet.
Note that the Ce-Ce interaction in CeB$_6$ has been extensively studied,
describing well the phase diagram under the magnetic field.
\cite{Shiina,Sakai,Shiba}
But 
the molecular field may change little and even stabilize the quadrupole
order. Therefore, we need not explicitly consider the molecular field
by regarding the crystal field as including the effect.
The charge anisotropy associated with 
the quadrupole order modulates the $5d$ states
through the intra-atomic $5d$-$4f$ Coulomb interaction.

We calculate the RXS intensity within the $E_1$ transition.
We take account of the above two processes, direct and indirect ones, 
of modulating the $5d$ states.
Both processes give rise to the RXS intensities 
on the $(00\frac{\ell}{2})$ and on the $(h0\frac{\ell}{2})$ spots.
Both give similar photon-energy dependences and the same azimuthal-angle 
dependence in agreement with the experiment.
However, the mechanism of direct modulation of the $5d$ band
gives rise to the intensities much larger than the mechanism
of indirect modulation through the $5d$-$4f$ Coulomb interaction
in a wide parameter range of the crystal field.
This suggests that the RXS intensities are mainly controlled 
by the lattice distortion.

This paper is organized as follows.
In \S~2, we analyze the buckling of sheets of B and C atoms.
In \S~3, we briefly summarize the formulae used in the calculation of
the RXS spectra.
In \S~4, we calculate the RXS spectra on two mechanisms.
Section 5 is devoted to concluding remarks. 

\section{Lattice Distortion} 

\begin{figure}[t]
\includegraphics[width=8.0cm]{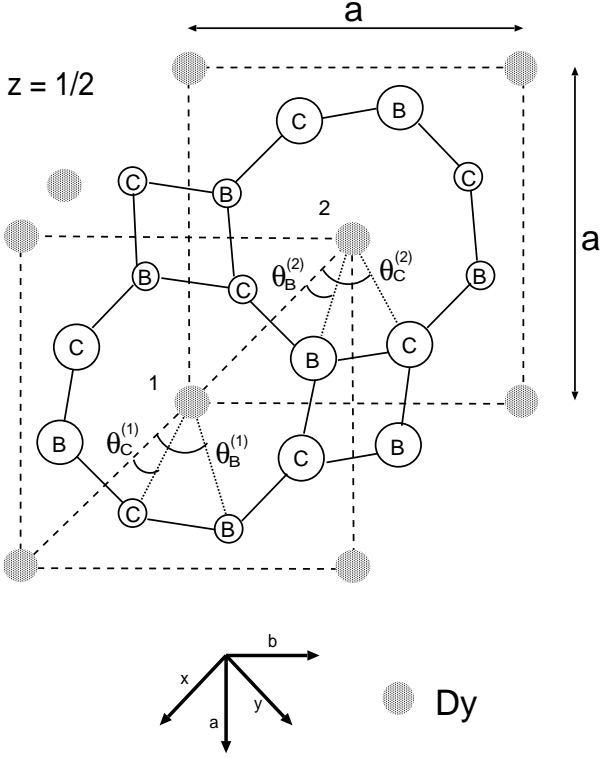}
\caption{
Sketch of a B$_2$C$_2$ sheet ($z=c/2$).
Open circles represent B and C atoms;
big and small circles move to positive and negative directions
along the $z$ axis, respectively.
The directions are reversed on the plane of $z=-c/2$.
Hatched circles represent Dy atoms at the plane of $z=0$.
\label{fig.distortion}}
\end{figure}
For making clear the effect of lattice distortion on electronic states,
we first calculate the electrostatic potential on the basis of a point
charge model. \cite{Hutchings}
Point charges $q_{Dy}$, $q_B$, and $q_C$ are placed 
on Dy, B, and C sites, respectively.
Figure \ref{fig.distortion} shows the positions of B and C atoms 
on the plane of $z=c/2$
and those of Dy atoms on the plane of $z=0$.
For the buckling of sheets of B and C atoms 
(up- and down-movements along the $c$ axis) 
specified in Fig.~\ref{fig.distortion}, 
the electrostatic potential is evaluated
within the second order of coordinates around Dy sites. 
As shown in Fig.~\ref{fig.cryst}(b), four inequivalent sites arises.
For sites $j$ ($=1\cdots 4$), the electrostatic potential is obtained as
\begin{equation}
 V_{\rm crys}(j) = A_2^0 Q_2^0
              + A_2^2(j) Q_2^2
              + A_{xy}(j)Q_{xy} ,
\label{eq.Vcry}
\end{equation}
with
\begin{align}
 Q_2^0 &= \frac{1}{2}(3z^2-r^2), \\
 Q_2^2 &= \frac{\sqrt{3}}{2}(x^2-y^2), \\
 Q_{xy}&= \sqrt{3}xy,
\end{align}
where the coefficients are given by
\begin{align}
A_{2}^{0} & = 
 \frac{2q_D}{a^3}
\left[ 1 -2 \frac{|q_B|}{q_D}
 \left(\frac{a}{R_{B}} \right)^3
\left\{ 1 - 3 \left( \frac{c}{2 R_{B}} \right)^2 \right\}
\right. \nonumber \\
&  \left.
         -2 \frac{|q_C|}{q_D}
  \left(\frac{a}{R_{C}} \right)^3
\left\{ 1 - 3 \left( \frac{c}{2 R_{C}} \right)^2 \right\} \right]
\nonumber \\
 &= \frac{2q_D}{a^3}\left[ 1 + 1.40\frac{|q_B|}{q_D} + 1.79\frac{|q_C|}{q_D}
  \right], \\
A_{2}^{2}(j) &= \Lambda_{B} \cos ( 2 \theta_{B}^{(j)} )
            + \Lambda_{C} \cos ( 2 \theta_{C}^{(j)} ), \label{eq.defA22} \\
A_{xy}(j) &=  \Lambda_{B} \sin ( 2 \theta_{B}^{(j)} )
          + \Lambda_{C} \sin ( 2 \theta_{C}^{(j)} ), \label{eq.defAxy}
\end{align}
with
\begin{align}
\Lambda_{B} & = - \frac{|q_B|}{a^3}
 \frac{60}{\sqrt{3}}  \left(\frac{r_B}{a} \right)^2
\left(\frac{a}{R_{B}}\right)^5
\left( \frac{c}{2 R_{B}} \right)^2 \frac{d_B}{\frac{c}{2}}
\nonumber \\
&= -22.3 \frac{|q_B|}{a^3} \frac{d_B}{\frac{c}{2}}, \\
\Lambda_{C} & = - \frac{|q_C|}{a^3}
 \frac{60}{\sqrt{3}}  \left(\frac{r_C}{a} \right)^2
\left(\frac{a}{R_{C}}\right)^5
\left( \frac{c}{2 R_{C}} \right)^2  \frac{d_C}{\frac{c}{2}}
\nonumber \\
&= -24.0 \frac{|q_C|}{a^3}  \frac{d_C}{\frac{c}{2}}.
\end{align}
The first term in eq. (\ref{eq.Vcry}) represents the crystal field
without lattice 
distortion, while the second and third terms arise from the buckling.
The $R_B$ ($=2.732$\AA) and $R_C$ ($=2.676$\AA) are distances from the origin 
to B and C sites, respectively. 
The $d_B$ and $d_C$ represent the absolute values of shifts 
along the $c$ axis from the $z=\pm c/2$ planes, respectively.
Angles $\theta_{B(C)}^{(j)}$ and $\theta_{B(C)}^{(j)}$
in eqs. (\ref{eq.defA22}) and (\ref{eq.defAxy}) are given by
\begin{align}
 \theta_B^{(1)} &= 65.6^{\circ}, \quad \theta_C^{(1)} = 19.1^{\circ},\nonumber\\
 \theta_B^{(2)} &= 180^{\circ}-65.6^{\circ}, 
   \quad \theta_C^{(2)} = 180^{\circ}-19.1^{\circ},\nonumber\\
 \theta_B^{(3)} &= 90^{\circ}+\theta_B^{(1)}, 
   \quad \theta_C^{(3)} = 90^{\circ}+\theta_C^{(1)},\nonumber\\
 \theta_B^{(4)} &= 90^{\circ}+\theta_B^{(2)}, 
   \quad \theta_C^{(4)} = 90^{\circ}+\theta_C^{(2)}.
\end{align}

Now we search for the local coordinate frames in which the third term 
in eq. (\ref{eq.Vcry}) is eliminated. Rotating the original 
coordinate frame by angle $\phi_j$ 
around the $c$ axis for each sublattice, we have the operators transformed as
\begin{align} 
 Q_2^2 &= \cos(2\phi_j)\tilde Q_2^2(j) - \sin(2\phi_j)\tilde Q_{xy}(j),
\nonumber\\
 Q_{xy} &= \sin(2\phi_j)\tilde Q_2^2(j) + \cos(2\phi_j)\tilde Q_{xy}(j),
\label{eq.rotQ}
\end{align}
where tilde operators $\tilde Q(j)$'s are represented with respect to 
the local coordinate frames.
$Q_2^0$ is unchanged. Inserting eq. (\ref{eq.rotQ}) into eq. (\ref{eq.Vcry}),
we have
\begin{equation}
 V_{crys}(j) = A_2^0 \tilde Q_2^0(j) + \tilde A_2^2(j)\tilde Q_2^2(j)
             + \tilde A_{xy}(j)\tilde Q_{xy}(j),
\end{equation}
with
\begin{align}
 \tilde A_{2}^{2}(j) &= \Lambda_{B} \cos [2(\theta_{B}^{(j)}-\phi_j)]
            + \Lambda_{C} \cos [ 2 (\theta_{C}^{(j)}-\phi_j)], \nonumber \\
\\
 \tilde A_{xy}(j) &= \Lambda_{B} \sin [2(\theta_{B}^{(j)}-\phi_j)]
          + \Lambda_{C} \sin [2(\theta_{C}^{(j)}-\phi_j)]. \nonumber \\
\end{align}
Condition $\tilde A_{xy}(j)=0$ determines $\phi_j$'s, which take values 
between $\theta_{B}^{(j)}$ and $\theta_{C}^{(j)}$.
For example, assuming $q_B=q_C=-(3/4)e$, $q_{Dy}=3e$ (e: proton charge),
$d_B=0.01$\AA, $d_C=0.02$\AA, we have $\phi_1=31.8^{\circ}$, 
$\phi_2=180^{\circ}-\phi_1$, 
$\phi_3=90^{\circ}+\phi_1$, $\phi_4=90^{\circ}+\phi_2$.
The principal axes thus estimated correspond well with the directions 
of the ordered magnetic moments in Phase III.\cite{Yamauchi}

The equivalent operator method allows us to write the crystal field energy
$H_{crys}(j)$ at site $j$ within the subspace of angular momentum $J$ as
\begin{equation}
 H_{\rm crys}(j) = D_{J}[3\tilde J_z^2(j)-J(J+1)]+E_{J}[\tilde J^2_x(j) 
                 - \tilde J^2_y(j)].
\label{eq.crystal}
\end{equation}

\leftline{\it (a) on the 5d bands}

The $5d$ states are forming an energy band with width $\sim 15$ eV 
through a hybridization with $s$ and $p$ states of neighboring B and C atoms
as well as $5d$ states of neighboring Dy atoms. 
We need the density of states (DOS) for calculating the RXS intensity.
We assume a Lorentzian with full width of half maximum 5 eV 
for the DOS's projected onto symmetries $xy$, $x^2-y^2$, $yz$, $zx$, 
and $3z^2-r^2$. The center of each component of the DOS 
is separate to each other in accordance with the first term of 
eq. (\ref{eq.crystal}), although the first term need not be respected so much
because of the large band effect.
Explicitly they are assumed to be
$(1/\pi)\Delta/((\epsilon-2.5)^2+\Delta^2)$ for $xy$, $x^2-y^2$,
$(1/\pi)\Delta/((\epsilon-7)^2+\Delta^2)$ for $yz$, $zx$,
and $(1/\pi)\Delta/((\epsilon-8.5)^2+\Delta^2)$ for $3z^2-r^2$, 
with energies in units of eV and $\Delta=2.5$ eV.
This arbitrary assumption for the DOS form may be justified by the fact that
the RXS spectra is not sensitive to the assumption.

The second term of eq. (\ref{eq.crystal}), which arises from the buckling 
of sheets of B and C atoms, gives rise to a small modification on the $5d$ band.
Although the actual modulation of the $5d$ states may come through 
the hybridization to the $s$ and $p$ states of B and C atoms, 
such effects can be included into the second term.
This term makes the local symmetry twofold.

\leftline{\it (b) on the 4f states}

Dy$^{3+}$ ion is approximately in the $4f^9$-configuration ($^6$H$_{15/2}$). 
Equation (\ref{eq.crystal}) is now applied to the subspace of $J=15/2$.
Since the $4f$ states are much localized than the $5d$ states,
the crystal field is much smaller here than that on the $5d$ states.
The coefficient $D_f$ of the first term is expected to be positive
from the analysis of magnetic susceptibility.\cite{Yamauchi} 
This leads to the lowest energy states $|\pm\frac{1}{2}\rangle$ 
and the next energy states $|\pm\frac{3}{2}\rangle$, both of which 
form Kramers' doublets ($|M\rangle$ represents the state of $J_z=M$).
The axial symmetry is kept instead of fourfold symmetry without the lattice
distortion. Terms of $O_4^4\equiv \frac{1}{2}(J_+^4 + J_-^4)$, which
admixes the states $|M\rangle$ with $|M\pm 4\rangle$, come from
the higher order expansion to make the local symmetry fourfold.
The detailed study along this line, however, is beyond the scope of 
the present study. 

In any event, the second term makes the local symmetry twofold.
The lowest energy state is admixed by $|M\rangle$ with $|M|>1/2$.
The quadrupole moment is ordered;
$\langle\tilde O_{x^2-y^2}\rangle$ 
($\equiv\frac{\sqrt{3}}{2}\langle\tilde J_x^2-\tilde J_y^2\rangle$)
takes a finite value with $\langle\tilde O_{xy}\rangle$ 
($\equiv\frac{\sqrt{3}}{2}\langle \tilde J_x\tilde J_y
+\tilde J_y\tilde J_x\rangle$)$=0$ in the local coordinate frame for
each sublattice ($\langle\cdots\rangle$ means the average over the lowest
Kramers doublet).
With increasing values of $|E_f|/D_f$, the average quadrupole moment
increases, as shown in Fig.~\ref{fig.parameter}(b).
It becomes largest, $\langle \tilde{O}_{x^2-y^2}\rangle=46.1$ at $D_{f}=0.1$.

\section{Cross Section of Resonant X-Ray Scattering}

\begin{figure}[t]
\includegraphics[width=8.0cm]{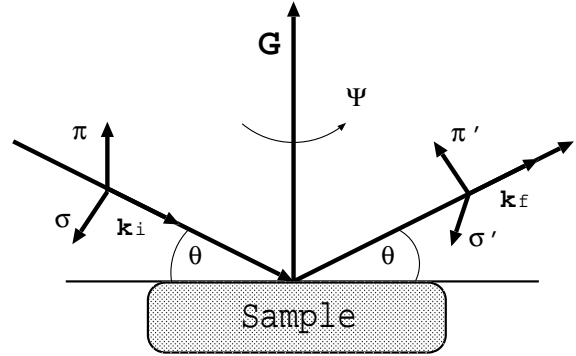}
\caption{
Scattering geometry. Incident photon with
wave vector ${\textbf k}_i$ and polarization $\sigma$ or $\pi$ 
is scattered into the state with wave vector ${\textbf k}_f$ and
polarization $\sigma'$ or $\pi'$ at Bragg angle $\theta$.
The sample crystal is rotated by azimuthal angle $\psi$
around the scattering vector ${\textbf G}={\textbf k}_f-{\textbf k}_i$.
\label{fig.geom}}
\end{figure}

We briefly summarize here the formulae used in the calculation of 
the RXS spectra in the next section.
The conventional RXS geometry is shown in Fig.~\ref{fig.geom};
photon with frequency $\omega$, momentum ${\textbf k}_i$ and polarization $\mu$ 
($=\sigma$ or $\pi$) is scattered into the state
with momentum ${\textbf k}_f$ and polarization $\mu'$ ($=\sigma'$ or $\pi'$). 
The scattering vector is defined as 
${\textbf G}\equiv{\textbf k}_f-{\textbf k}_i$.
Near the Dy $L_{\rm III}$ absorption edge, a $2p$ core electron is virtually
excited to $5d$ states in the $E_1$ process.
Subsequently it recombines with the core hole.
Since the $2p$ states are well localized around Dy sites,
it is a good approximation to describe the scattering tensor
as a sum of contributions from each site of the core hole.
Therefore, the cross section in the $E_1$ process is given by
\begin{equation}
 I_{\mu\to\mu'}({\textbf G},\omega) \propto
 | \sum_{\alpha\alpha'}P'^{\mu'}_{\alpha}
    M_{\alpha\alpha'}({\textbf G},\omega)P^{\mu}_{\alpha'}
 |^2,
\label{eq.cross}
\end{equation}
where 
\begin{equation}
 M_{\alpha\alpha'}({\textbf G},\omega) = \frac{1}{\sqrt{N}}
  \sum_j M_{\alpha\alpha'}(j,\omega) \exp(-i{\textbf G}\cdot{\textbf r}_j),
\label{eq.scatensor}
\end{equation}
with
\begin{equation}
 M_{\alpha\alpha'}(j,\omega) = 
  \sum_{\Lambda}
  \frac{\langle\psi_n|x_\alpha(j)|\Lambda\rangle
  \langle \Lambda|x_{\alpha'}(j)|\psi_j\rangle}
       {\hbar\omega-(E_{\Lambda}-E_j)+i\Gamma},
\label{eq.dipole}
\end{equation}
\begin{table}[t]
\caption{Geometrical factors}
\begin{tabular}{cccc}
$\alpha$&$(P^\sigma)_\alpha$&$(P'^{\sigma'})_\alpha$&$(P'^{\pi'})_\alpha$\\ 
\hline
  1     & $\cos\beta\cos\psi$ & $\cos\beta\cos\psi$ 
        & $-\sin\theta\cos\beta\sin\psi+\cos\theta\sin\beta$\\
  2     & $- \sin\psi$        & $- \sin\psi$        
        & $-\sin\theta\cos\psi+\cos\theta\sin\beta$\\
  3     & $-\sin\beta\cos\psi$& $-\sin\beta\cos\psi$
        & $\cos\theta\cos\beta$\\
\end{tabular}
\label{tab.azim}
\end{table}
where $N$ is the number of Dy sites.
Note that the cross section is an order of $N$.
The $P^\mu$ and $P'^{\mu}$ are geometrical factors for the incident 
and scattered photons, respectively. Their explicit forms 
are given in Table \ref{tab.azim}.
The $|\psi_j\rangle$ represents the initial state with energy $E_j$.
The intermediate state $|\Lambda\rangle$ consists of an excited electron
on $5d$ states and a hole on $2p$ states with energy $E_{\Lambda}$.
The $\Gamma$ is the life-time broadening width of the core hole.
The dipole operators $x_\alpha(j)$'s are defined as
$x_1(j)=x$, $x_2(j)=y$, and $x_3(j)=z$ in the coordinate frame fixed 
to the crystal axes with the origin located at the center of site $j$.
The scattering amplitude $M_{\alpha\alpha'}({\textbf G},\omega)$
contains the square of the dipole matrix element 
$A_{dp}=\langle 5d|r|2p\rangle =
   \int_0^{\infty} R_{5d}(r)rR_{2p}(r)r^2{\rm d}r $
with $R_{5d}(r)$ and $R_{2p}(r)$ being the radial wavefunctions 
for the $5d$ and $2p$ states.
For Dy$^{3+}$ atom, it is estimated as $2.97\times 10^{-11}$ cm 
in the $4f^9$ configuration
within the HF approximation.\cite{Cowan}

\section{Calculation of RXS spectra} 

In the $E_1$ transition, an electron is excited from $2p$ states to $5d$ 
states at a Dy site. A $2p$ core hole is left behind, and its state
is split into the states of $j_p=3/2$ and $j_p=1/2$ ($j_p$ is the total 
angular momentum) due to the strong spin-orbit interaction. 
We consider only the $j_p=3/2$ states ($L_{\rm III}$ edge).
We describe the photoexcited $5d$ electron in the band
by introducing a local Green's function,
\begin{equation}
 G^{5d}_{m^d}(\hbar\omega)
   = \int_{\epsilon_F}^{\infty} 
   \frac{\rho^{5d}_{m^d}(\epsilon)}{\hbar\omega-\epsilon+i\delta}
   {\rm d}{\epsilon},
\label{eq.5dgreen}
\end{equation}
where $\rho^{5d}_{m^d}(\epsilon)$ is the $m^d$ component of the $d$ DOS
defined in \S 2.
The Fermi energy $\epsilon_F$ is set to be zero 
so that the $5d$ band is almost empty.

Finite RXS intensities on superlattice spots arise from modulating
the $5d$ states with wave vectors of superlattice spots.
There are two origins to giving rise to such modulation. 
One is a direct influence from the buckling of sheets of B and C atoms, 
which is represented by the second term of eq. (\ref{eq.crystal}).
Another is the charge anisotropy of the $4f$ states in the quadrupole 
ordering phase. We discuss both origins separately.
In the actual calculation, we specify the local coordinate frames
with $\phi_1=28^{\circ}$, $\phi_2=180^{\circ}-28^{\circ}$, 
$\phi_3=90^{\circ}+28^{\circ}$, $\phi_4=90^{\circ}+152^{\circ}$ 
for four sublattices in accordance with the experiment.\cite{Yamauchi}

\subsection{Direct influence of lattice distortion}

Let the $E_1$ transition take place at a particular site (called as ``origin"). 
The excited $5d$ electron is attracted 
by the core hole potential at the origin.
What is more important is that the $5d$ electron is under the influence
of the second term of eq. (\ref{eq.crystal}).
Taking account of the multiple scattering from these terms at the origin,
we evaluate the resolvent $1/(\hbar\omega-H_{\rm int})$ with respect to
the intermediate-state Hamiltonian $H_{\rm int}$:
\begin{align}
&  \left(\frac{1}{\hbar\omega - H_{\rm int} + i\Gamma}\right)_
{m^ds^d\lambda;m'^ds'^d\lambda'}
\nonumber \\
   = &[G^{5d}_{m^d}(\hbar\omega+i\Gamma-\epsilon_\lambda)^{-1}
     \delta_{\lambda\lambda'}\delta_{m^dm'^d}\delta_{s^ds'^d} 
  \nonumber\\ 
     &- V_{m^ds^d\lambda;m'^ds'^d\lambda'}]^{-1},
\label{eq.matrix1}
\end{align}
where $m^d$ and $s^d$ specify the orbital and spin of the $d$ electron,
respectively. The $\epsilon_{\lambda}$ represents the energy of the core hole
with $\lambda$ in the $j_p=3/2$ subspace.  The scattering potential 
$V_{m^ds^d\lambda;m'^ds'^d\lambda'}$ includes the second term of
eq. (\ref{eq.crystal}) and the Coulomb interaction between the $5d$ electron 
and the $2p$ hole.
The latter quantity is expressed in terms of the Slater integrals,
which are evaluated within the HF approximation in a Dy$^{3+}$ atom
(see Table \ref{tab.slater}).\cite{Com1}
The core hole life-time width is set to be $\Gamma=2.5$ eV.
Equation (\ref{eq.matrix1}) is numerically evaluated.
\begin{table}[t]
\caption{Slater integrals and the spin-orbit interaction
for Dy$^{3+}$ atoms in
the Hartree-Fock approximation (in units of eV).}
\begin{center}
\begin{tabular}{llll}
\hline
\hline
 $F^{k}(4f,4f)$    & $F^{k}(2p,5d)$  & $F^{k}(2p,4f)$  & $F^{k}(4f,5d)$  \\
 $F^0$ \hspace*{0.2cm} 32.19 & 
 $F^0$ \hspace*{0.2cm} 16.13 & 
 $F^0$ \hspace*{0.2cm} 44.62 & 
 $F^0$ \hspace*{0.2cm} 14.70 \\
 $F^2$ \hspace*{0.2cm} 15.31 & 
 $F^2$ \hspace*{0.2cm} 0.489 & 
 $F^2$ \hspace*{0.2cm} 1.982 & 
 $F^2$ \hspace*{0.2cm} 3.614 \\
 $F^4$ \hspace*{0.2cm} 9.607 &  
                & 
                & 
 $F^4$ \hspace*{0.2cm} 1.741 \\
 $F^6$ \hspace*{0.2cm} 6.911 & 
                & 
                &                 \\
\hline
                  & 
 $G^{k}(2p,5d)$  & 
 $G^{k}(2p,4f)$  & 
 $G^{k}(4f,5d)$  \\
                  & 
 $G^1$ \hspace*{0.2cm} 0.414 & 
 $G^2$ \hspace*{0.2cm} 0.207 & 
 $G^1$ \hspace*{0.2cm} 1.615 \\
                  & 
 $G^3$ \hspace*{0.2cm} 0.245 & 
 $G^4$ \hspace*{0.2cm} 0.133 & 
 $G^3$ \hspace*{0.2cm} 1.321 \\
                  &
                  &  
                  & 
 $G^5$ \hspace*{0.2cm} 1.009 \\
\hline
$\zeta_{4f}=$ 0.273 & $\zeta_{5d}=$ 0.181 &  & \\
\hline
\hline
\end{tabular}
\end{center}
$\ast$In the RXS calculation, the above values of the anisotropic terms
are reduced by multiplying a factor 0.8, while the values for
$F^{(0)}(nl,n'l')$ are replaced by much smaller values,
$F^{(0)}(4f,5d) = 3.0$, $F^{(0)}(4f,4f) = 7.0$,
$F^{(0)}(2p,5d) = 4.0$, $F^{(0)}(2p,4f) = 12.0$.
\label{tab.slater}
\end{table}

\begin{figure}[t]
\includegraphics[width=8.0cm]{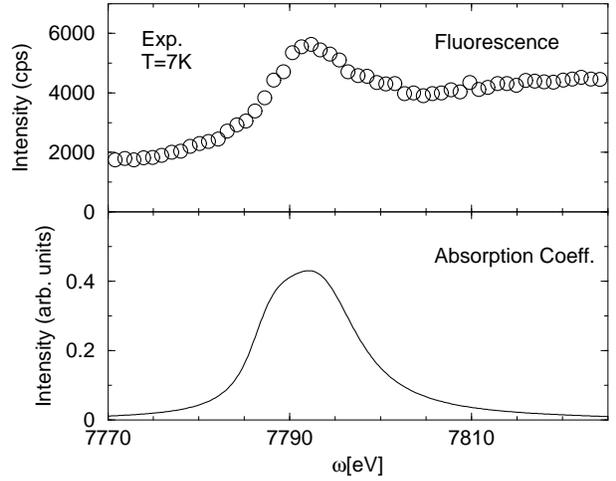}
\caption{
Absorption coefficient $A(\omega)$ (lower panel) in comparison with 
the $L_{\rm III}$-edge fluorescence spectra (upper panel).$^{17)}$
\label{fig.absorp}}
\end{figure}

Before calculating the RXS spectra, we touch on 
the absorption coefficient. We calculate the absorption coefficient 
$A(\omega)$ in the $E_1$ process from the resolvent by using the relation,
\begin{align}
 A(\omega) & \propto  \sum_{j} \sum_{\alpha}
   \langle\psi_j|x^\alpha(j)|m^ds^d\lambda\rangle 
\nonumber \\
&  \times
   \left(-\frac{1}{\pi}\right)
 {\rm Im}\left(\frac{1}{\hbar\omega - H_{\rm int} + i\delta}\right)_
    {m^ds^d\lambda;m'^ds'^d\lambda'} \nonumber\\
  & \times\langle m'^ds'^d\lambda'|x^\alpha(j)|\psi_j\rangle .
\label{eq.absorp}
\end{align}
Here ${\rm Im}X$ indicates the imaginary part of the quantity $X$.
Figure \ref{fig.absorp} shows the calculated $A(\omega)$ 
in comparison with the fluorescence experiment.\cite{Hirota}
We adjust the core hole energy such that the calculated peak 
coincides with the experimental one. 
The calculated curve reproduces the experimental one.
Since the spectra is proportional to the $d$ DOS 
when the $5d$-$2p$ Coulomb interaction is neglected, 
the assumed DOS seems reasonable.
The intensity seen in the high energy region of the experiment may come from
the $d$ symmetric states mixing with $3s$, $3p$ states of B and C atoms,
which is outside our interest.

Now we discuss the RXS spectra. The resolvent above calculated is used to 
calculate the scattering amplitude at the origin with the help of the relation,
\begin{align}
&  \sum_{\Lambda}\frac{|\Lambda\rangle\langle\Lambda|}
       {\hbar\omega-(E_{\Lambda}-E_j)+i\Gamma} \nonumber \\
  &= \sum_{m^ds^d\lambda}\sum_{m'^ds'^d\lambda'}
|m^ds^d\lambda\rangle \nonumber \\
& \times 
 \left(\frac{1}{\hbar\omega - H_{\rm int} + i\Gamma}\right)_
{m^ds^d\lambda;m'^ds'^d\lambda'}
  \langle m'^ds'^d\lambda'| .
\label{eq.intermed}
\end{align}
This expression is independent of the quadrupole ordering
$4f$ states.  It is inserted into eq. (\ref{eq.dipole})
to calculate an RXS amplitude.
The extension to general site $j$ is straightforward.
In the coordinate frame fixed to crystal 
(not in the local coordinate frames), their forms are given by
\begin{align}
\hat M(1,\omega) &=
\left( \begin{array}{ccc}
          \xi(\omega)  & \eta(\omega)  & 0 \\
          \eta(\omega) & \zeta(\omega) & 0 \\
          0            & 0             & \gamma(\omega)
       \end{array}
\right), \nonumber \\ 
\hat M(2,\omega) &=
\left( \begin{array}{ccc}
          \zeta(\omega)  & \eta(\omega) & 0 \\
          \eta(\omega)   & \xi(\omega)  & 0 \\
          0              & 0            & \gamma(\omega)
       \end{array}
\right), \nonumber \\ 
\hat M(3,\omega) &=
\left( \begin{array}{ccc}
          \zeta(\omega)   & -\delta(\omega) & 0 \\
          -\delta(\omega) & \xi(\omega)     & 0 \\
          0               & 0               & \gamma(\omega)
       \end{array}
\right), \nonumber \\ 
\hat M(4,\omega) &=
\left( \begin{array}{ccc}
          \xi(\omega)     & -\delta(\omega) & 0 \\
          -\delta(\omega) & \zeta(\omega)   & 0 \\
          0               & 0               & \gamma(\omega)
       \end{array}
\right). \label{eq.Mamp4}
\end{align}

\leftline{\it (a) ${\textbf G}=(00\frac{\ell}{2})$ with $\ell$ odd integers.}

Owing to the factor $\exp(-i{\textbf G}\cdot{\textbf r}_j)$ 
in eq. (\ref{eq.scatensor}), 
the total amplitude is given by a combination of 
$\hat M(1,\omega)+\hat M(2,\omega)-\hat M(3,\omega)-\hat M(4,\omega)$.
Thus, we have  the final form, 
\begin{equation}
\frac{\hat M({\textbf G},\omega)}{\sqrt{N}} =
\left( \begin{array}{ccc}
          0              & \eta(\omega)   & 0 \\
          \eta(\omega)   & 0              & 0 \\
          0              & 0              & 0
       \end{array}
\right).
\label{eq.final1}
\end{equation}
The geometrical factors are given by setting $\beta=0$ 
in Table \ref{tab.azim}, which are combined to
eq. (\ref{eq.final1}) to calculate the scattering intensity.
We have the RXS intensity as a function of  azimuthal angle $\psi$ as
\begin{align}
 I_{\sigma\to\sigma'}({\textbf G},\omega) &\propto |\eta(\omega)|^2\sin^2 2\psi,
\nonumber \\
 I_{\sigma\to\pi'}({\textbf G},\omega) &\propto
   |\eta(\omega)|^2\sin^2\theta\cos^2 2\psi,
\label{eq.azim1}
\end{align}
with $\theta$ the Bragg angle.
Here $\psi$ is defined such that $\psi=0$ corresponds to
the scattering plane containing the $b$ axis.

\begin{figure}[t]
\includegraphics[width=8.0cm]{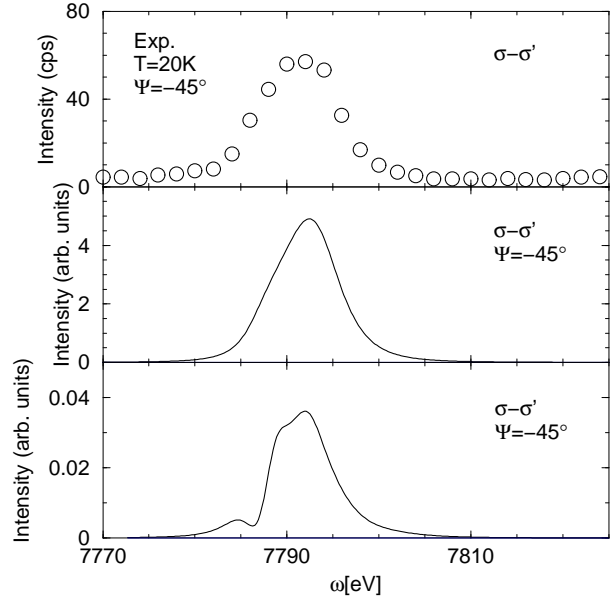}
\caption{
RXS spectra for ${\textbf G}=(00\frac{5}{2})$ at $\psi=-45^\circ$
in the $\sigma\to\sigma'$ channel, as a function of photon energy.
Top: experimental spectra at $T=20$ K (Phase II).$^{17)}$
Middle: Calculated spectra by taking account of the
direct influence of lattice distortion
with $E_d=-0.1$ eV.
Bottom: 
Calculated spectra by taking account of the
influence of quadrupole ordering $4f$ states 
with $E_f/D_f=-0.2$.
\label{fig.spec}
}
\end{figure}

Figure \ref{fig.spec} shows the calculated RXS spectra as a function of 
photon energy in comparison with the 
experiment ($\psi=-45^\circ$).\cite{Hirota}
The crystal field parameter is set to be $E_d=-0.1$ eV 
in eq. (\ref{eq.crystal}).
As shown in the middle panel, the calculated spectra show
a single-peak in agreement with the experiment.
(Only the $\sigma\to\sigma'$ channel gives finite intensity 
for $\psi=-45^\circ$.)
The photon energy dependence in the $\sigma\to\pi'$ channel
is found to be the same as in the $\sigma\to\sigma'$ channel
in the calculation.  On the other hand, 
an extra peak has been observed in the $\sigma\to\pi'$ channel
at $\hbar\omega=7782$ eV (pre-edge peak) for $\psi=0$.\cite{Hirota}
This peak may come from the electric quadrupole ($E_2$) transition.

\begin{figure}[h]
\includegraphics[width=8.0cm]{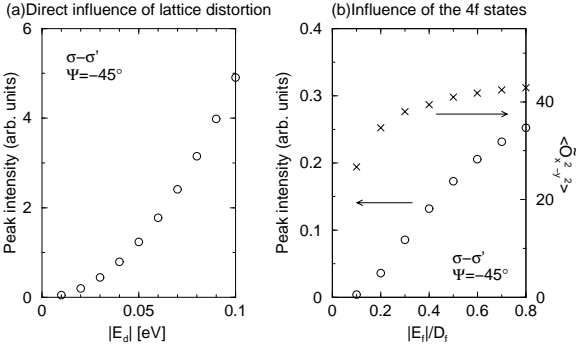}
\caption{
(a) Main peak intensity as a function of the crystal field parameter $|E_d|$
on the $5d$ states, which is calculated by taking account of the
direct influence of lattice distortion.
(b) Main peak intensity as a function of the crystal field parameter 
$|E_f|/D_f$ on the $4f$ states, which is calculated by taking account of the
influence of quadrupole ordering $4f$ states (open circles).
Crosses represent the quadrupole moments $\langle \tilde{O}_{x^2-y^2} \rangle$
in the local coordinate frame for each sublattice.
\label{fig.parameter}}
\end{figure}

Figure \ref{fig.parameter}(a) plots the peak intensity as a function of $|E_d|$ 
(at $\psi=-45^\circ$ in the $\sigma\to\sigma'$ channel).
The intensity of the ``main" peak increases with increasing values of $|E_d|$.
It is nearly proportional to $|E_d|^2$.

\begin{figure}[t]
\includegraphics[width=8.0cm]{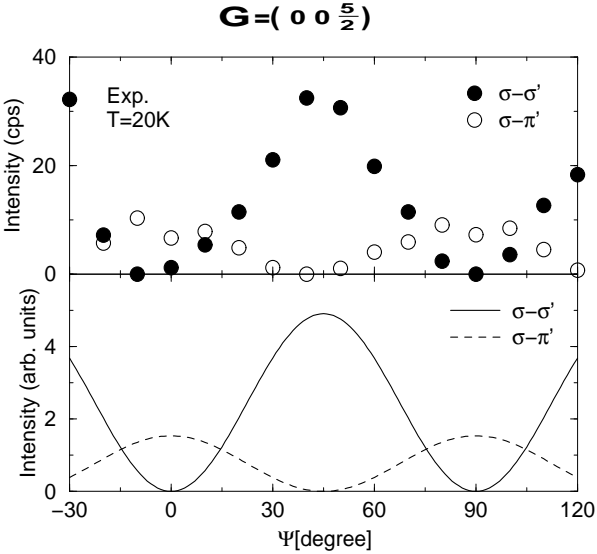}
\caption{
Azimuthal angle dependence of the RXS intensity of the main peak
on ${\textbf G}=(00\frac{5}{2})$,
in comparison with the experiment at $T=20$ K.$^{17)}$
\label{fig.azim1}}
\end{figure}

Figure \ref{fig.azim1} shows the azimuthal angle dependence of 
the main peak intensity for ${\textbf G}=(00\frac{5}{2})$, in good agreement 
with the experiment.
The same dependence as eq. (\ref{eq.azim1}) has been proposed
on the basis of the symmetry of the scattering tensor.\cite{Matsumura}
Note that the intensity ratio between
the $\sigma\to\sigma'$ channel and the $\sigma\to\pi'$ channel is
determined by a geometrical factor;
the oscillation amplitude of intensity in the $\sigma\to\pi'$ channel 
is the factor $\sin^2\theta=0.312$ smaller than that 
in the $\sigma\to\sigma'$ channel.

\leftline{\it (b) ${\textbf G}=(h0\frac{\ell}{2})$ with $h$ and $\ell$ odd integers.}

\begin{figure}[t]
\includegraphics[width=8.0cm]{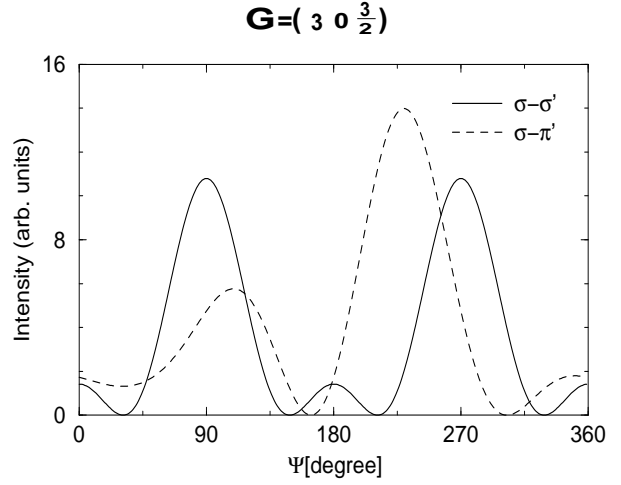}
\caption{
Azimuthal angle dependence of the RXS intensity of the main peak
on ${\textbf G}=(30\frac{3}{2})$.
\label{fig.azim2}}
\end{figure}

The total amplitude is given by a combination of 
$\hat M(1,\omega)-\hat M(2,\omega)-\hat M(3,\omega)+\hat M(4,\omega)$.
Thus, we have  the final form,
\begin{equation}
\frac{\hat M({\textbf G},\omega)}{\sqrt{N}} =
\left( \begin{array}{ccc}
 \frac{1}{2}(\xi(\omega)-\zeta(\omega)) & 0 & 0 \\
 0 & \frac{1}{2}(\zeta(\omega)-\xi(\omega)) & 0 \\
          0              & 0                    & 0
       \end{array}
\right).
\label{eq.final2}
\end{equation}
Combining the geometrical factor in Table I to eq. (\ref{eq.final2}),
we obtain the scattering intensity.
The spectral shape as a function of photon energy is found almost the same as
that for the $(00\frac{\ell}{2})$ spot, so that we omit the corresponding
figure. The azimuthal angle dependence is given by
\begin{align}
 I_{\sigma\to\sigma'}({\textbf G},\omega) &\propto
    \frac{1}{4}|\xi(\omega)-\zeta(\omega)|^2
      (\cos^2\beta\cos^2\psi-\sin^2\psi)^2,
\nonumber \\
 I_{\sigma\to\pi'}({\textbf G},\omega) &\propto
    \frac{1}{4}|\xi(\omega)-\zeta(\omega)|^2 \nonumber\\
  &\times [\cos\theta\sin\beta(\cos\beta\cos\psi+\sin\psi)
\nonumber \\
& 
   -\sin\theta(1+\cos^2\beta)\sin\psi\cos\psi)]^2,
\label{eq.azim2}
\end{align}
where $\beta$ is determined from $\tan\beta=(2hc)/(\ell a)$.
Figure \ref{fig.azim2} shows the azimuthal dependence of the peak intensity
for the $(30\frac{3}{2})$ spot.
A large non-resonant intensity has been observed in the 
$\sigma \rightarrow \sigma'$ channel, and the resonant behavior is not
clear in the experiment.\cite{Tanaka,Hirota}
The non-resonant intensity may come from the
Thomson scattering due to the lattice distortion.
We hope that the resonant behavior discussed here is observed in
the $\sigma \rightarrow \pi'$ channel, since the non-resonant intensity
is expected to disappear in this channel.

\subsection{Influence of quadrupole ordering 4f states}

The initial state is evaluated in \S~2, where
the $4f$ quadrupole moment is ordered. 
The intermediate state is evaluated by the following steps.
Let the $E_1$ transition take place at the origin. 
The complex of $4f$ electrons and the $2p$ hole is assigned as
the eigenstate $|\nu\rangle$ with energy $E_{\nu}$,
which is calculated by diagonalizing the matrix of the intra-atomic
Coulomb interaction.
To keep the matrix size manageable, the space of $4f$ states is restricted
within the space of $J=15/2$.
This restriction causes only minor errors in the RXS spectra,
since the RXS amplitude contains the overlap between the $4f$ states 
in the intermediate state and that in the initial state, which becomes
very small for the $4f$ states outside the $J=15/2$ subspace.
Since the photoexcited $5d$ electron interacts with the $2p$ hole and 
with $4f$ electrons,
the complex of $4f$ electrons and the $2p$ hole serves as a scatterer to
the $5d$ electron. Thus we have the resolvent at the origin,
\begin{align}
 &  \left(\frac{1}{\hbar\omega - H_{\rm int} + i\Gamma}\right)_
{m^ds^d\nu;m'^ds'^d\nu'}
\nonumber \\
 &  = [G^{5d}_{m^d}(\hbar\omega+i\Gamma-E_\nu)^{-1}
     \delta_{\nu\nu'}\delta_{m^dm'^d}\delta_{s^ds'^d} 
  \nonumber\\ 
     &  - U_{m^ds^d\nu;m'^ds'^d\nu'}]^{-1},
\label{eq.matrix2}
\end{align}
where $m^d$ and $s^d$ specify $5d$ states.
The potential $U_{m^ds^d\nu;m'^ds'^d\nu'}$ includes the $5d$-$4f$ Coulomb
interaction as well as the $5d$-$2p$ Coulomb interaction.
They are expressed in terms of the Slater integrals, which are given
in Table II.\cite{Cowan,Com1}
Since the $5d$-$4f$ Coulomb interaction is implicitly included in
the $5d$-band energy, we eliminate the average of the $5d$-$4f$ interaction 
from the potential $U$. 
The energies of the $4f$ states coming from the crystal field 
much smaller than other energies in the intermediate state,
and thus can be neglected.
We do not include the crystal field on the $5d$
states due to the lattice distortion (the second term of 
eq. (\ref{eq.crystal})).
Within the present approximation,
the right hand side of eq. (\ref{eq.matrix2}) becomes a matrix 
with dimensions $640\times 640$ ($640=10\times 64$), 
which are numerically inverted.

The resolvent thus obtained is the same on four sublattices.
The scattering amplitudes become different on different sublattices
after the transition matrix elements from the initial state are taken
into account. 
The scattering amplitude is found to take the same forms as 
eqs. (\ref{eq.Mamp4}).
Therefore, the azimuthal-angle dependence is the same as eqs. (\ref{eq.azim1})
and (\ref{eq.azim2}). As regards the photon-energy dependence,
the bottom panel in Fig.~\ref{fig.spec} shows 
the RXS spectra at $\psi=-45^{\circ}$ in the $\sigma\to\sigma'$ channel
for ${\textbf G}=(00\frac{5}{2})$.
We put $E_f/D_f=-0.2$.
The spectral shape is not so different from the curve given in 
the preceding subsection, although a small hump appears at the low energy
side.  Difference is the magnitude of the intensity. 
Figure \ref{fig.parameter}(b) shows the main-peak intensity
as a function of $|E_f|/D_f$.
It increases with increasing values of $|E_f|/D_f$.
In the same figure, we have also plotted the sublattice quadrupole moment as
a function of $|E_{f}|/D_{f}$.  It also increases with increasing values
of $|E_{f}|/D_{f}$.
This coincidence is plausible, since the charge anisotropy in the $4f$ states
increases with increasing the ordered quadrupole moment.
However, its magnitude remains much smaller than those given by the direct
influence of lattice distortion
in a wide parameter range of crystal field by
comparison with the curve in Fig.~\ref{fig.parameter}(a).

\section{Concluding Remarks}

We have studied the mechanism of RXS at the $L_{\rm III}$ edge 
in the quadrupole ordering phase of DyB$_2$C$_2$.
Having analyzed the effect of the bucking of sheets of B and C atoms
on the $5d$ and $4f$ states, we have constructed an effective model
that the crystal field is acting on the $5d$ and $4f$ states
with the principal axes different for different sublattices.
We have calculated the RXS spectra in the $E_1$ process
by treating the $5d$ states as a band and the $4f$ states as localized states. 
We have considered two mechanisms separately that
the lattice distortion directly modulates the $5d$ band 
and that the charge anisotropy of the quadrupole ordering
$4f$ states modulate the $5d$ band through the $5d$-$4f$ intra-atomic 
Coulomb interaction.
We have found that both mechanisms give rise to the RXS intensities on
$(00\frac{\ell}{2})$ and $(h0\frac{\ell}{2})$ spots
with similar photon-energy dependences and the same 
azimuthal angle dependence.
Both explain well the experimental RXS spectra.
However, it is shown that
the former mechanism gives rise to the intensity
much larger than the latter one for a wide parameter range  of crystal field. 
This suggests that the main-peak of the RXS spectra is  
not a direct reflection of the quadrupole order but
mainly controlled by the lattice distortion.
To confirm this observation more quantitatively, 
band structure calculations may be useful since
the $5d$ states are considerably extended in space.
This study is left in the future.

As regards the pre-edge peak, we have estimated its 
intensity within the $E_2$ transition.  In that estimate, we have used
the same initial state as discussed above and have taken account
of the full multiplets of the $f^{10}$-configuration
for the intermediate state.
The transition matrix element has been evaluated by the atomic Hartree-Fock 
wave function.\cite{Cowan}
The pre-edge peak intensity thus evaluated is found to be more than 
three-order of magnitude smaller than the main-peak intensity 
evaluated by the mechanism of the charge anisotropy of the 
quadrupole ordering $4f$ states. 
This is inconsistent with the experiments, where the pre-edge
peak intensity is the same order of magnitude to the main peak intensity.
Clarifying this point is also left in the future.

\section*{Acknowledgments}

We would like to thank S. W. Lovesey, Y. Tanaka, and T. Inami 
for valuable discussions.
This work was partially supported by 
a Grant-in-Aid for Scientific Research 
from the Ministry of Education, Science, Sports and Culture.


\def\vol(#1,#2,#3){{\textbf #1} (#2) #3}



\begin{thebibliography}{99}

\bibitem{Murakami98a} Y. Murakami, H. Kawata, M. Tanaka, T. Arima, Y. Moritomo
and Y. Tokura:
                      Phys. Rev. Lett. \vol(80,1998,1932). 

\bibitem{Murakami98b} Y. Murakami, J. P. Hill, D. Gibbs, M. Blume, I. Koyama,
                      M. Tanaka, H. Kawata, T. Arima, Y. Tokura,
                      K. Hirota and Y. Endoh:
                      Phys. Rev. Lett. \vol(81,1998,582). 

\bibitem{Murakami99c} M. von Zimmermann, J.P. Hill, D. Gibbs, M. Blume, 
                      D. Casa, B. Keimer, Y. Murakami, Y. Tomioka 
                      and Y. Tokura:
                      Phys. Rev. Lett. \vol(83,1999,4872). 

\bibitem{Murakami00b} M. Noguchi, A. Nakazawa, T. Arima, Y. Wakabayashi, 
                      H. Nakao and Y. Murakami:
                      Phys. Rev. B \vol(62,2000,R9271).    

\bibitem{Ishihara1} S. Ishihara and S. Maekawa:
                   Phys. Rev. Lett. \vol(80,1998,3799).

\bibitem{Elfimov} I. S. Elfimov, V. I. Anisimov and G. Sawatzky:
                  Phys. Rev. Lett. \vol(82,1999,4264).

\bibitem{Benfatto} M. Benfatto, Y. Joly and C. R. Natoli:
                   Phys. Rev. Lett. \vol(83,1999,636).

\bibitem{Takahashi1} M. Takahashi, J. Igarashi and P. Fulde:
                    J. Phys. Soc. Jpn. \vol(68,1999,2530).

\bibitem{Takahashi2} M. Takahashi, J. Igarashi and P. Fulde:
                    J. Phys. Soc. Jpn. \vol(69,2000,1614).

\bibitem{Nakao01} H. Nakao, K. Magishi, Y. Wakabayashi, Y. Murakami,
                  K. Koyama, K. Hirota, Y. Endoh and S. Kunii:
                  J. Phys. Soc. Jpn. \vol(70,2001,1857).

\bibitem{Nagao} T. Nagao and J. Igarashi:
                J. Phys. Soc. Jpn. \vol(70,2001,2892).  

\bibitem{Igarashi} J. Igarashi and T. Nagao:
                   J. Phys. Soc. Jpn.  \vol(71,2002,1771). 

\bibitem{Shiina} R. Shiina, H. Shiba and P. Thalmeier:
                J. Phys. Soc. Jpn. \vol(66,1997,1741).

\bibitem{Sakai} O. Sakai, R. Shiina,  H. Shiba and P. Thalmeier:
                J. Phys. Soc. Jpn. \vol(66,1997,3005).

\bibitem{Shiba} H. Shiba, O. Sakai and R. Shiina:
                J. Phys. Soc. Jpn. \vol(68,1999,1988).

\bibitem{Tanaka} Y. Tanaka, T. Inami, T. Nakamura, H. Yamauchi,
                 H. Onodera, K. Ohoyama and Y. Yamaguchi:
                 J. Phys. Condens. Matter \vol(11,1999,L505).

\bibitem{Hirota} K. Hirota, N. Oumi, T. Matsumura, H. Nakao,
                 Y. Wakabayashi, Y. Murakami and Y. Endoh:
                 Phys. Rev. Lett. \vol(84,2000,2706).

\bibitem{Matsumura} T. Matsumura, N. Oumi, K. Hirota, H. Nakao,
                    Y. Murakami, Y. Wakabayashi, T. Arima, S. Ishihara
                    and Y. Endoh:
                    Phys. Rev. B \vol(65,2002,094420).

\bibitem{Yamauchi} H. Yamauchi, H. Onodera, K. Ohoyama, T. Onimaru,
                   M. Kosaka, M. Ohashi and Y. Yamaguchi:
                   J. Phys. Soc. Jpn. \vol(68,1999,2057).

\bibitem{Adachi} H. Adachi, H. Kawata, M. Mizumaki, T. Akao, M. Sato,
                 N. Ikeda, Y. Tanaka and H. Miwa:
                 Phys. Rev. Lett. \vol(89,2002,206401).

\bibitem{Lovesey} S. W. Lovesey and K. S. Knight:
                  Phys. Rev. B \vol(64,2001,094401).

\bibitem{Hutchings} M. T. Hutchings: Solid State Physics,
                    \vol(16,1964,227).

\bibitem{Cowan} R. Cowan: {\emph The Theory of Atomic Structure and Spectra}
                (University of California Press, Berkeley, 1981).

\bibitem{Com1} The anisotropic terms of the Coulomb interaction
are slightly reduced in solids; we use the atomic values 
in Table \ref{tab.slater} by reducing them with multiplying a factor 0.8.
On the other hand, the values of $F^0(n\ell,n'\ell')$
are considerably screened in solids,
so that they are replaced by much smaller values.

\end{thebibliography}
\end{document}